\documentclass[12pt]{main}

\title{First principles and scanning tunneling spectroscopical evidences for thermodynamically stable "on-top" sulfur divacancy in monolayer WS$_{2}$}

\author[1*]{Weiru Chen}
\author[1,2,3*]{John C. Thomas}
\author[1]{Yihuang Xiong}
\author[4,5]{Zhuohang Yu}
\author[6]{Da Zhou}
\author[4,5]{Shalini Kumari}
\author[3]{Zhongwei Dai}
\author[5,6,7, 8]{Joshua A. Robinson}
\author[5,6,7,8]{Mauricio Terrones}
\author[3,4]{Archana Raja}
\author[2,3]{Sin\'ead Griffin}
\author[2,3]{Alexander Weber-Bargioni}
\author[1,9,10$\dagger$]{Geoffroy Hautier}

\affil[1]{Thayer School of Engineering, Dartmouth College, 14 Engineering Dr, Hanover, NH 03755, United States}
\affil[2]{Materials Sciences Division, Lawrence Berkeley National Laboratory, Berkeley, CA 94720, United States}
\affil[3]{Molecular Foundry, Lawrence Berkeley National Laboratory, Berkeley, CA 94720, United States}
\affil[4]{Kavli Energy NanoScience Institute, University of California Berkeley,
Berkeley, CA 94720, United States}

\affil[5]{Department of Materials Science and Engineering, The Pennsylvania State University, University Park, PA 16082, United States}
\affil[6]{Center for Two-Dimensional and Layered Materials, The Pennsylvania State University, University Park, PA, 16802, United States}
\affil[7]{Department of Physics, The Pennsylvania State University, University Park, PA, 16802, United States}
\affil[8]{Department of Chemistry, The Pennsylvania State University, University Park, PA, 16802, United States}
\affil[9]{Department of Materials Science and NanoEngineering, Rice University, Houston, TX, 77005, United States}
\affil[10]{Rice Advanced Materials Institute, Rice University, Houston, TX, 77005, United States}
\affil[$\dagger$]{geoffroy.hautier@rice.edu}
\affil[*]{These authors contributed equally.}

\usepackage{amsmath} 
\usepackage{ulem}
\newcommand{\strike}[1]{\bgroup\markoverwith{\textcolor{black}{\rule[0.5ex]{2pt}{0.4pt}}}\ULon{#1}}

\date{}

\begin{document}
\maketitle

\begin{abstract}
Chalcogen vacancies in monolayer transition metal dichalcogenides (TMDs), such as WS$_{2}$, play a crucial role in various applications ranging from optoelectronics and catalysis to quantum information science (QIS), making their identification and control essential. This study focuses on WS$_{2}$  single vacancy and vacancy pairs. Using first principles computations, we investigate their thermodynamic stabilities and electronic structures. We identify an "on-top" divacancy configuration where two vacancies sit on top of each other to be the only energetically stable complex with a binding energy of 160 meV. We compute a small difference in electronic structure with a shift of the unoccupied state by 140 meV for the divacancy complex and observe electronic state shift during Scanning Tunneling Spectroscopy of a series of vacancy in WS$_2$ providing spectroscopical evidence for the presence of this defect.

\end{abstract}

\maketitle

Transition metal dichalcogenides (TMDs) MX$_{2}$ (M = transition metal atoms: W, Mo; X: chalcogen atoms: S, Se) are an interesting class of semiconducting two dimensional (2D) materials providing a wide array of rich material properties that include opto-electronic transitions, bandgap tunability, Fermi level modification, and dopant amenability \cite{novoselov2005two,  wang2012electronics, butler2013progress, xu2013graphene}. As a monolayer, 1H-phase TMDs exhibit semiconducting characteristics with direct band gaps in the visible range at the $K$ point of the Brillouin zone\cite{mak2010atomically, zeng2013optical, rasmussen2015computational}. Their intrinsic quantum confinement nature, reduced dielectric screening, as well as strong spin-orbit coupling (SOC) have given rise to various technologically-attractive phenomenons including strong electron-hole interactions\cite{mak2013tightly, ross2013electrical}, catalytic activities\cite{lauritsen2007location, zong2008enhancement} and high carrier mobility\cite{radisavljevic2011single, cui2015high}. As many semiconductors, TMDs are sensitive to structural defects, making their identification and control of great importance \cite{lin2016defect}. In particular, chalcogen vacancies are arguably the most common defects and have been shown to influence various TMDs properties. Chalcogen vacancies have been shown to control catalysis, carrier transport or optical emission\cite{ouyang2016activating, tsai2017electrochemical,qiu2013hopping,tongay2013defects, carozo2017optical}. In quantum applications, chalcogen vacancies can act as single photon emitters in TMDs \cite{srivastava2015optically, schuler2020electrically, mitterreiter2021role, hotger2023spin}, a key component towards quantum information science applications\cite{weber2010quantum, dreyer2018first, zhang2020material, wolfowicz2021quantum}. While many earlier studies have focused on single chalcogen vacancies, there is a growing interest in understanding vacancy complexes \cite{sun2024unveiling}.The proximity of individual vacancies within a complex influences the effective electronic structure through dipole-dipole coupling, potential wavefunciton overlap, as spin-spin interaction, ultimately determining their functionalities as potential quantum emitters or catalysts. Additionally, the geometric arrangement of vacancy complexes can create novel potential landscapes, leading to the breaking of symmetries. Understanding and controlling defect complexes provides new opportunities to tailor defect-induced functionalities of 2D semiconductors. In this paper, we study the interactions of sulfur vacancy pairs in monolayer WS$_{2}$. By using first-principles computations, we examine the thermodynamic stability and electronic structures of sulfur vacancy pairs and identify an energetically favorable divacancy configuration where two vacancies sit on top of each other. Experimental scanning tunneling microscopy (STM) and scanning tunneling spectroscopy (STS) further provide evidence for the formation of such divacancy complex defects.

\begin{figure}[h!]
    \centering
    \includegraphics[width = \textwidth]{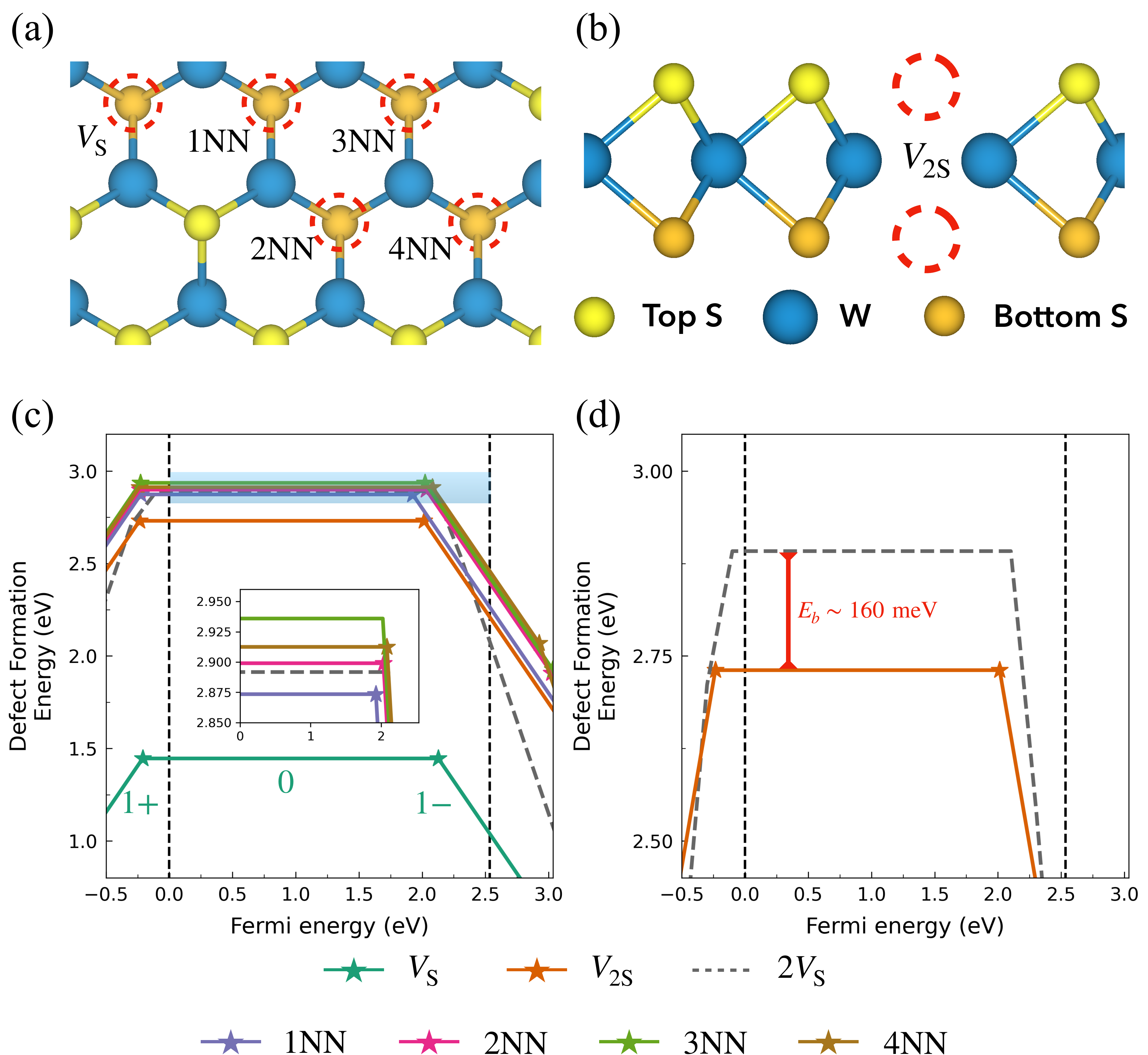}
    \caption{Simulated defect configurations and thermodynamic stability and binding energy of relevant defects. (a): sulfur single vacancy $V_{\mathrm{S}}$ and up to 4$^{\mathrm{th}}$ NN  vacancy complexes ($V_{\mathrm{S}}-V_{\mathrm{S}}$): 1NN, 2NN, 3NN and 4NN, separated by 3.16\AA, 5.48\AA, 6.32\AA, and 8.37\AA, respectively. (b): sulfur divacancy configuration $V_{\mathrm{2S}}$ where two vacancies sit on top of each other. (c): formation energy of sulfur single vacancy $V_{\mathrm{S}}$, divacancy $V_{\mathrm{2S}}$, and nearest neighbor complexes 1NN to 4NN from charge state 2- to 2+ evaluated under sulfur poor condition. Gray dashed line indicating two infinitely separated vacancies $2V_{\mathrm{S}}$ competing with the formation of vacancy paries. Inset shows zoomed-in energy range marked in the shaded area. (d): a binding energy around 160 meV is observed by the divacancy configuration at neutral charge state. Band edge positions are marked with vertical dashed lines. Fermi energy is referenced to the VBM.}
    \label{fig:main_fig_1}
\end{figure}

The thermodynamic stability of these defects was computed through defect formation energy using the PBE0 hybrid functional (see Methodology in Supporting Information). Figure \ref{fig:main_fig_1}c shows the formation energy vs Fermi level plots for sulfur single and vacancy pairs at charge state 2-, 1-, 0, 1+ and 2+ evaluated under sulfur poor conditions. 2- and 2+ states are not stable within the bandgap. The sulfur rich conditions can be found in Figure S1. We compute a vacancy complex binding energy for all these pairs, see Figure S2). We observe that the divacancy configuration (with two vacancy on top of each other (see Figure \ref{fig:main_fig_1}b) exhibits a favorable binding energy $E_{b} \sim 160$ meV at the neutral charge state. All other vacancy complexes have negative or significantly smaller binding energy. Figure \ref{fig:main_fig_1}d illustrates this by comparing the formation energy of the divacancy and two infinitely separated vacancies. We note that this binding energy observed for the divacancy is largely unaffected if we use a different amount of exact exchange in the hybrid functionals (see Figure S2b). The value of binding energy is relatively small compared to complex defects in other semiconductors which can be of a few eVs \cite{chen2010intrinsic,varley2016defects,lee2021stability}. This low binding energy indicates that the divacancy might not be very favorable at high synthesis temperatures. They could however form during low temperature annealing or cooling down after synthesis. We note that kinetic effects might be important and prevent their formation despite their favorable driving force at lower temperatures. In any case, we show that there is a thermodynamic driving force at low temperature for forming these "on-top" divacancies and that they might be difficult to avoid in certain synthesis conditions. Recent work has measured the emission properties of single vacancy and vacancy complexes in WS$_2$ highlighting that only nearest-neighbor divacancy provide bright and stable optical emission\cite{sun2024unveiling}. We note that the favorable binding energy could make the formation of these optically unfavorable "on-top" divacancy difficult to bypass and could be an important challenge to address when designing future optimal single-photon emission systems based on vacancy in WS$_2$. 

\begin{figure}[h!]
    \centering
    \includegraphics[width=0.7\textwidth]{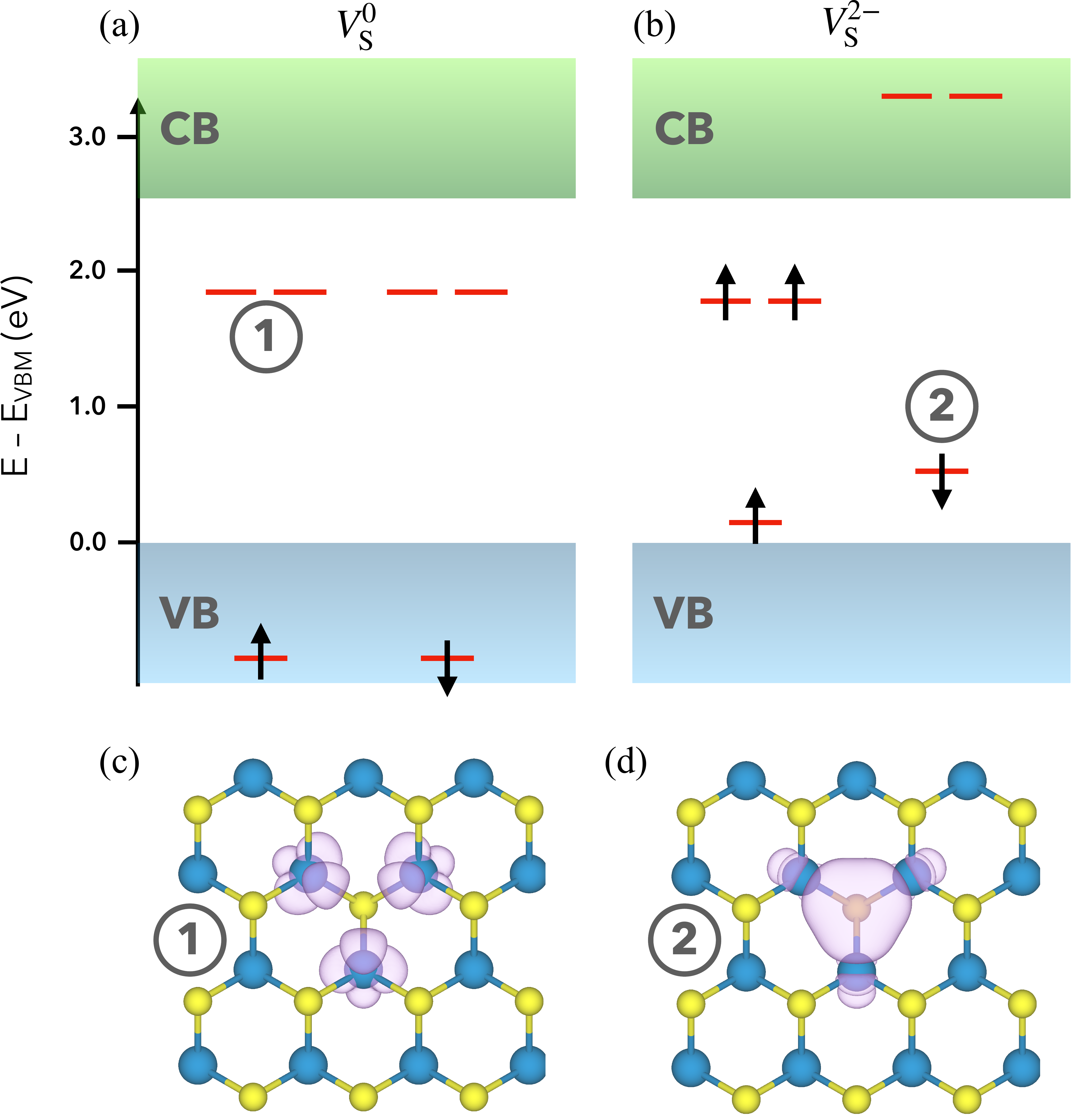}
    \caption{Single-particle Kohn-Sham energy levels for single sulfur vacancy at (a) 0 and (b) 2- state. The occupied levels at 0 charge state are buried deep within the valence band. SOC is not applied to defect states. (c) and (d), charge density distributions for $e$ and $a_{1}$ states labeled in (a) and (b), respectively. Isocontour value: 0.002 \AA$^{-3}$.}
    \label{fig:V_S_electronic_structure}
\end{figure}

\begin{figure}[h!]
    \centering
    \includegraphics[width = 1\textwidth]{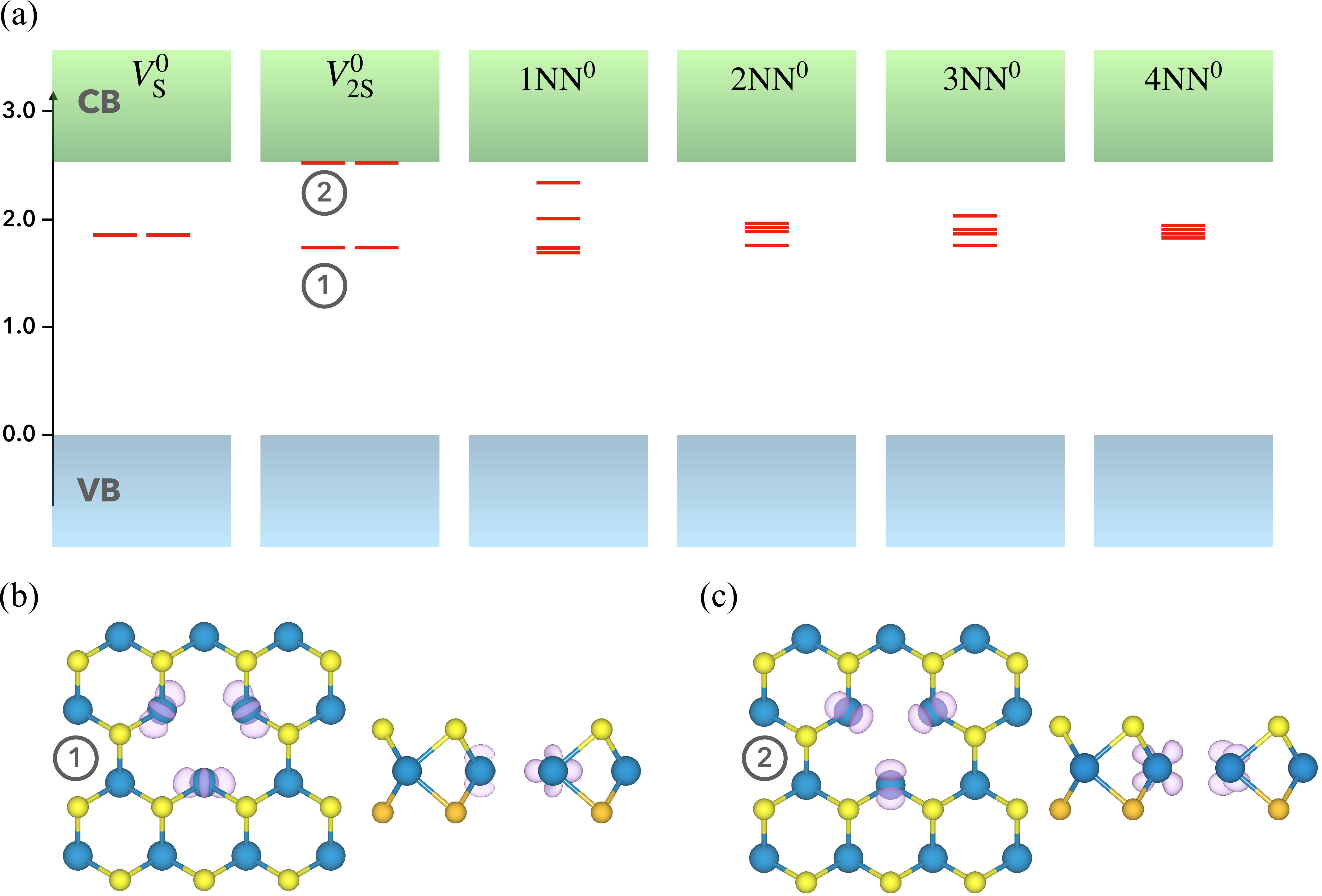}
    \caption{Electronic structure for sulfur single vacancy, divacancy and nearest neighbor vacancy complexes at neutral charge state. Only one spin channel is shown for simplicity. SOC is not applied to defect levels. (b) and (c), Top and side view of the charge density distributions of the lower and higher defect states labeled by 1 and 2 in (a), respectively. Isocontour value: 0.004 \AA$^{-3}$.}
    \label{fig:main_fig_3}
\end{figure}

\begin{figure}[h!]
    \centering
    \includegraphics[width = 0.7\textwidth]{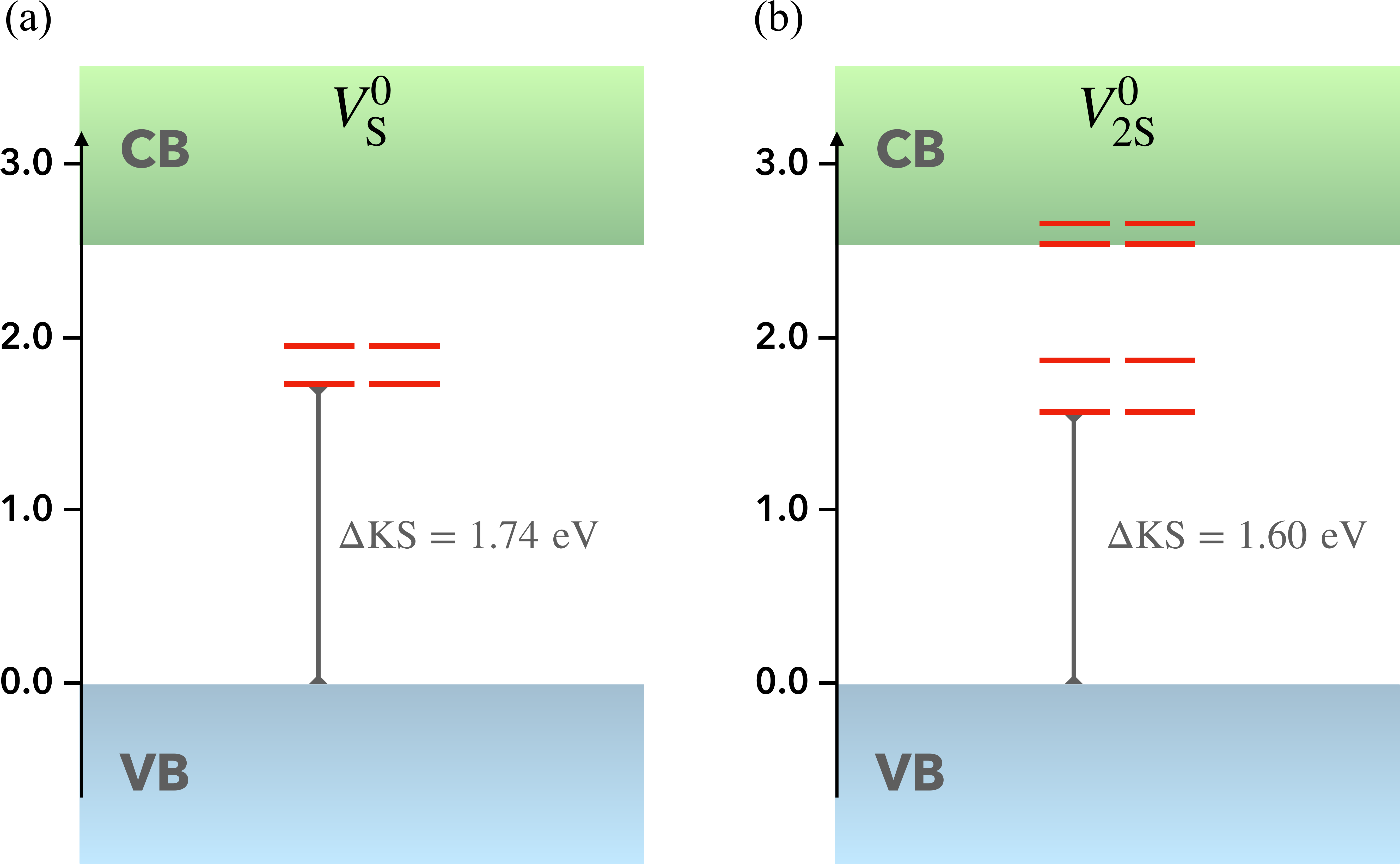}
    \caption{Single particle Kohn-Sham energy levels for (a) single and (b) divacancy at neutral charge state including SOC. An electronic transition from the occupied VBM to an unoccupied defect state is described by the Kohn-Sham energy difference $\Delta \mathrm{KS}$.}
    \label{fig:main_fig_4}
\end{figure}

In the electronic structure analysis of these vacancy complexes, the sulfur single vacancy possesses a $C_{3v}$ symmetry and hosts a singlet ground state at neutral charge state with two unoccupied degenerate $e$ states and a fully occupied $a_{1}$ state resonant with valence band. This $a_{1}$ state can be observed clearly if we look at $2-$ charge state (see Figure \ref{fig:V_S_electronic_structure}). 
The charge density distribution of the $e$ state exhibits  $d_{x^2 - y^2}$ and $d_{xy}$ orbitals from the three nearby tungsten atoms while the $a_{1}$ state mixes $d_{z^2}$ orbitals.
Similar electronic structures of single neutral chalcogen vacancy have been observed in previous studies in WS$_2$ and other TMDs\cite{liu2013sulfur, carozo2017optical, lu2018passivating, naik2018substrate, schuler2019large, schuler2020electrically, mitterreiter2021role, tsai2022antisite, zhao2023electrical}. From now on, we will only show defect states that are within the bandgap in the paper for simplicity. When we introduce one more sulfur vacancy, the electronic structures of the NN vacancy complexes converges as expected towards that of isolated vacancy with increasing separation. The proximity of another vacancy splits the unoccupied levels by 0.74, 0.66, 0.18, 0.27 and 0.1 eV for divacancy, 1NN, 2NN, 3NN and 4NN, respectively, as shown in Figure \ref{fig:main_fig_3}a.   
The divacancy configuration, shown to have a favorable binding energy,
exhibits a unique signature in its electronic structure. Figure \ref{fig:main_fig_3}a shows the ground state electronic structures of divacancy and the wavefunctions associated with defect states (Figure \ref{fig:main_fig_3}b-c). The divacancy configuration has the $D_{3h}$ symmetry and hosts a singlet ground state at neutral charge state. Interestingly, it shows two unoccupied degenerate in-gap states as the single vacancy and additional two unoccupied degenerate states resonant with the conduction band. The wavefunctions of these states indicate that they are localized and originated from nearby tungsten $d$ orbitals. Similar electronic structure of the divacancy configuration has been reported previously\cite{singh2021atypical}. Further incorporation of SOC splits the in-gap states degeneracy by 0.2 eV and 0.3 eV for single and divacancy configuration, respectively, resulting in four doubly degenerate in-gap states for both single and divacancy configuration. The defect states resonant with the conduction band for the divacancy configuration moves further into the conduction band (see Figure \ref{fig:main_fig_4}).
We observe that the unoccupied defect state is shifted towards the VBM in the case of the divacancy by 140 meV with a Kohn-Sham energy difference ($\Delta \rm{KS}$) between VBM and defect state of 1.74 eV and 1.60 eV for single and divacancy configuration, respectively. As photoluminescence measurements have been commonly used to study and detect chalcogen vacancies in TMDs, we can wonder if the divacancy could have a detectable optical signature. Unfortunately, currently in low temperature (77 K) photoluminescence measurements, single sulfur vacancy usually show a full width at half maximum (FDHM) from 100 to 130 meV\cite{carozo2017optical,bianchi2024engineering}. This would make it challenging to clearly observe the divacancy signature within the current sample quality. However, scanning tunneling spectroscopy (STS) can detect such a divacancy.

\begin{figure}[h!]
    \centering
    \includegraphics[width = 1\textwidth]{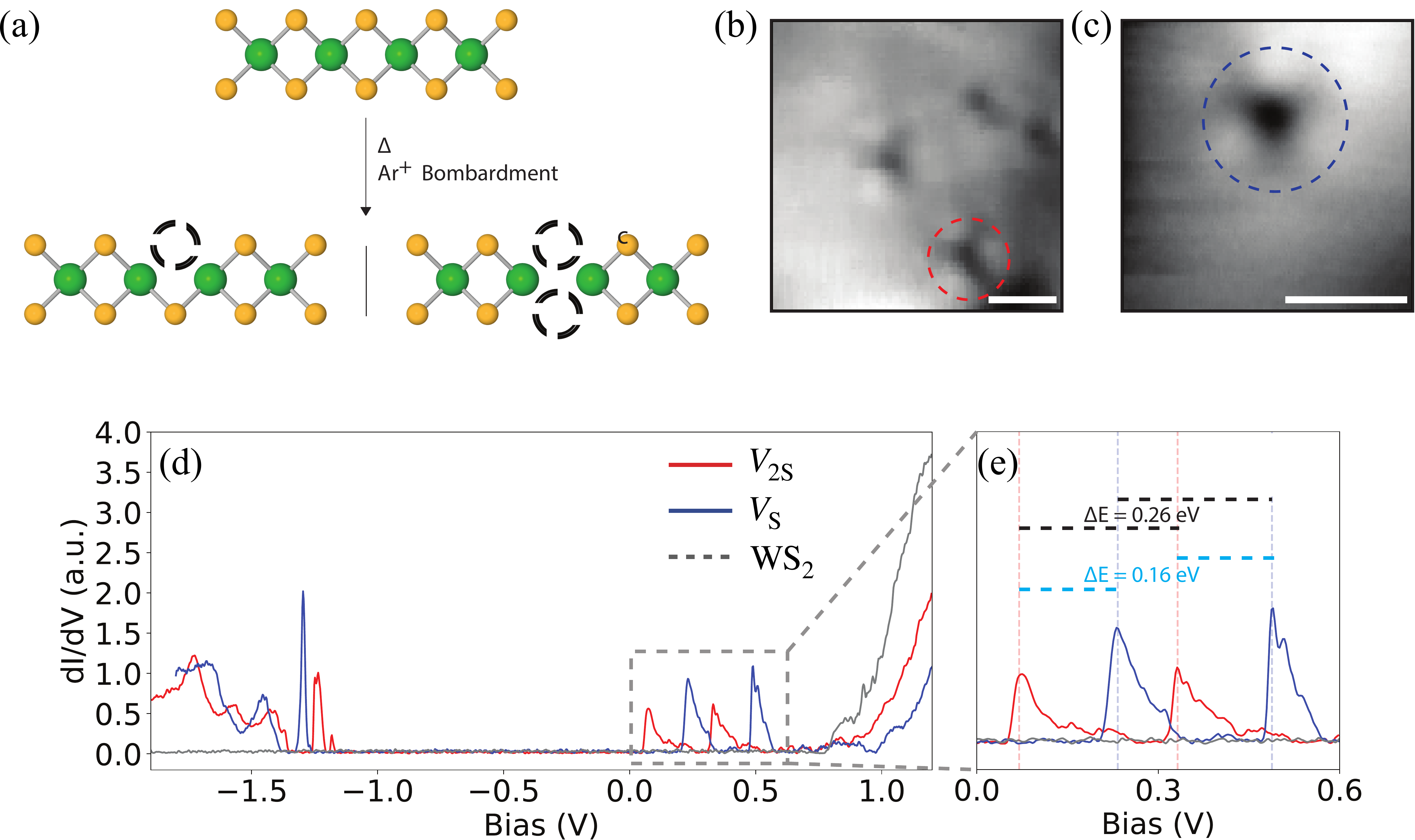}
    \caption{Divacancy scanning tunneling spectroscopy. (a): A schematic of our procedure to prepare defects within WS$_2$ is depicted. (b) and (c): Two defects are depicted that are indistinguishable in scanning tunneling microscopy ($I_{tunnel}$ = 30 pA, $V_{sample}$ = $1.2$ V). Scale bars, 1 nm. (d), Scanning tunneling spectroscopy ($V_{modulation}$ = 5 mV) of each defect is shown (red, (b); blue, (c)), where an  energetic shift of 0.16 eV is found ((d) and (e))  between both defects that otherwise exhibit matched spectroscopic signatures. We attribute this shift to the neutral divacancy $V_{\mathrm{2S}}^{0}$ formation.}
    \label{fig:main_fig_5}
\end{figure}

The computed unoccupied state shift for single and divacancy configurations suggests that we should be able to detect the divacancy signature with STM, in combination with STS\cite{PhysRevLett.123.076801,Thomas2022}. We purposefully introduce defects in our sample through sputtering and annealing producing a tunable amount of chalcogen vacancies\cite{rossi2025graphene,Thomas2024}. The sample is heated during Ar$^{+}$ bombardment (Figure \ref{fig:main_fig_5}a), where rearrangement and formation of stable defects occurs as the sample is further annealed. Two distinct V$_{\rm S}$ signatures can be identified with STS (Figure \ref{fig:main_fig_5}d) that are indistinguishable under conventional STM imaging, as shown in Figure \ref{fig:main_fig_5} b-c. Both signatures are characteristic of the sulfur vacancy defect with a split of unoccupied state of 0.26 eV (see Figure \ref{fig:main_fig_5} d-e). However, one of the spectrum is downshifted by 0.16 eV from the other, where the isolated VS electronic structure tungsten d orbital peak locations are consistent with
literature values\cite{Thomas2024}. This agrees with the computed downshift of the divacancy (see Figure \ref{fig:main_fig_4}) which is around 0.14 eV. It is thus likely that the shifted STS signal is related to the "on-top" divacancy complex. 

In summary, we used first-principles computations to study the thermodynamic stability and electronic structure of sulfur divacancy and nearest neighbor vacancy complexes in monolayer WS$_{2}$. We identified the "on-top" divacancy configuration be the only thermodynamically stable vacancy complex with a binding energy of 160 meV. We noted a shifted electronic transition from the occupied valence band maximum to the unoccupied defect state within the band gap for the "on-top" divacancy configuration compared to single vacancy. STS measurements pointed out to spectroscopical signature consistent with the "on-top" divacancy in WS$_2$. As quantum applications are considering sulfur vacancy and their complex as single-photon emitters in WS$_2$, the favorable binding energy of the "on-top" divacancy could have important implications on the generation and control of defect emitters in WS$_2$.

This work has been supported by the U.S. Department of Energy, Office of Science, Basic Energy Sciences in Quantum Information Science under Award Number DE-SC0022289. This research used resources of the National Energy Research Scientific Computing Center, a DOE Office of Science User Facility supported by the Office of Science of the U.S. Department of Energy under Contract No. DE-AC02-05CH11231 using NERSC award BES-ERCAP0020966.

\bibliographystyle{ieeetr} 
\bibliography{main}
\end{document}


\maketitle

\section{Methodology}
\subsection{Computational details}
All computations were performed using Vienna Ab-initio Simulation Package (\textsc{VASP})\cite{G.Kresse-PRB96,G.Kresse-CMS96} and the projector-augmented wave (PAW) method\cite{P.E.Blochl-PRB94} with the global hybrid functional PBE0 \cite{perdew1996rationale, adamo1999toward}. Each charged defect was simulated at the dilute limit using a 144 atom orthorhombic monolayer supercell within a 24 \AA vacuum layer. A plane-wave basis cutoff energy of 400 eV was used and the Brillouin zone was sampled using the $\Gamma$ point only. Defect structure relaxations were performed at fixed volume until the force on each iron is less than 0.01 eV/\AA. The formation energy of each charged defect was computed using \textsc{PyCDT}\cite{broberg2018pycdt}. For a general defect $X$ at charge state $q$, its formation energy is given by:
$E^{\rm{f}}\left[X^{q}\right]$ is given by:
\begin{equation}
    E^{\rm{f}}\left[X^{q}\right] = E_{\rm{tot}}\left[X^{q}\right] - E_{\rm{tot}}^{\rm{bulk}} - \sum n_{i}\mu_{i} + qE_{F} + E_{\rm{corr}},
    \label{eq:formation_energy}
\end{equation}
where the first two terms are the total energies of the supercell containing the defect and the pristine one, respectively. The third term takes care of the energy exchange with the chemical reservoirs when creating the defect with $n_{i}$ being the number of atoms with species $i$ and corresponding chemical potential $\mu_{i}$ being added ($n_{i}>0 $) or removed ($n_{i}<0$). The fourth term accounts for the electron exchange with the host material controlled by the Fermi level $E_{F}$, which is conventionally referenced to the valence band maximum (VBM). The last one is the correctional term for the finite-size effect, for which we employed methods proposed by Komsa et al.\cite{komsa2013finite,komsa2014charged} as implemented in \textsc{SLABCC}\cite{tabriz2019slabcc}. We evaluated the defect formation energy under sulfur-rich conditions ($\Delta \mu_{\mathrm{W}}=-2.53$ eV ) and sulfur-poor conditions  ($\Delta \mu_{\mathrm{S}} = -1.26$ eV). The chemical potentials of each species are referenced to their elemental forms.

The band edge positions of pristine WS$_{2}$ was computed by incorporating 22\% of Fock exchange with a dense $k$-point mesh of 12$\times$ 12 $\times$ 1. Defect levels were computed using 7\% Fock exchange. This scheme to tune the amount of Fock exchange individually for band edge positions and defect levels is recently proposed for a proper description of the bandgap while reasonably fulfilling the generalized Koopmans' condition on defect levels\cite{chen2022nonunique}. Spin-orbit-coupling is taken into account throughout for the electronic structures unless otherwise specified. Both charge corrections and vacuum level alignment are included for single-particle Kohn-Sham energy levels.
\subsection{Experimental preparation and measurement}
Monolayer islands of WS$_2$ were grown on graphene/SiC substrates with an ambient pressure CVD approach. A graphene/SiC substrate with 10 mg of WO$_3$ powder on top was placed at the center of a quartz tube, and 400 mg of sulfur powder was placed upstream. The furnace was heated to 900 $^{\circ}$C and the sulfur powder was heated to 250 $^{\circ}$C using a heating belt during synthesis. A carrier gas for process throughput was used (Ar gas at 100 sccm) and the growth time was 60 min. The CVD grown WS$_{2}$/MLG/SiC was further annealed \emph{in vacuo} at 400 $^{\circ}$C for 2 hours. 

WS$_2$ was sputtered with an argon ion gun (SPECS, IQE 11/35) that operated at 0.1 keV energy with 60$^{\circ}$ off-normal incidence at a pressure of 5$\times$10$^{-6}$ mbar and held at 600 $^{\circ}$C. A measure of current (0.6$\times$10$^{-6}$ A) enabled the argon ion flux to be estimated at (1.5$\times$10$^{13}$ \begin{math}\frac{ions}{cm^2s}\end{math}), where the sample was irradiated for up to 10 seconds.

All measurements were performed with a Createc GmbH scanning probe microscope operating under ultrahigh vacuum (pressure $<$ 2$\times$10$^{-10}$~mbar) at liquid helium temperatures ($T<6$~K). Either etched tungsten or platinum iridium tips were used during acquisition. Tip apexes were further shaped by indentations onto a gold substrate for subsequent measurements taken over a defective substrate. STM images are taken in constant-current mode with a bias applied to the sample. STS measurements were recorded using a lock-in amplifier with a resonance frequency of 683 Hz and a modulation amplitude of 5 mV. 
\section{Figure S1-S2}
\begin{figure*}[h!]
    \centering
    \includegraphics[width = 1.0 \textwidth]{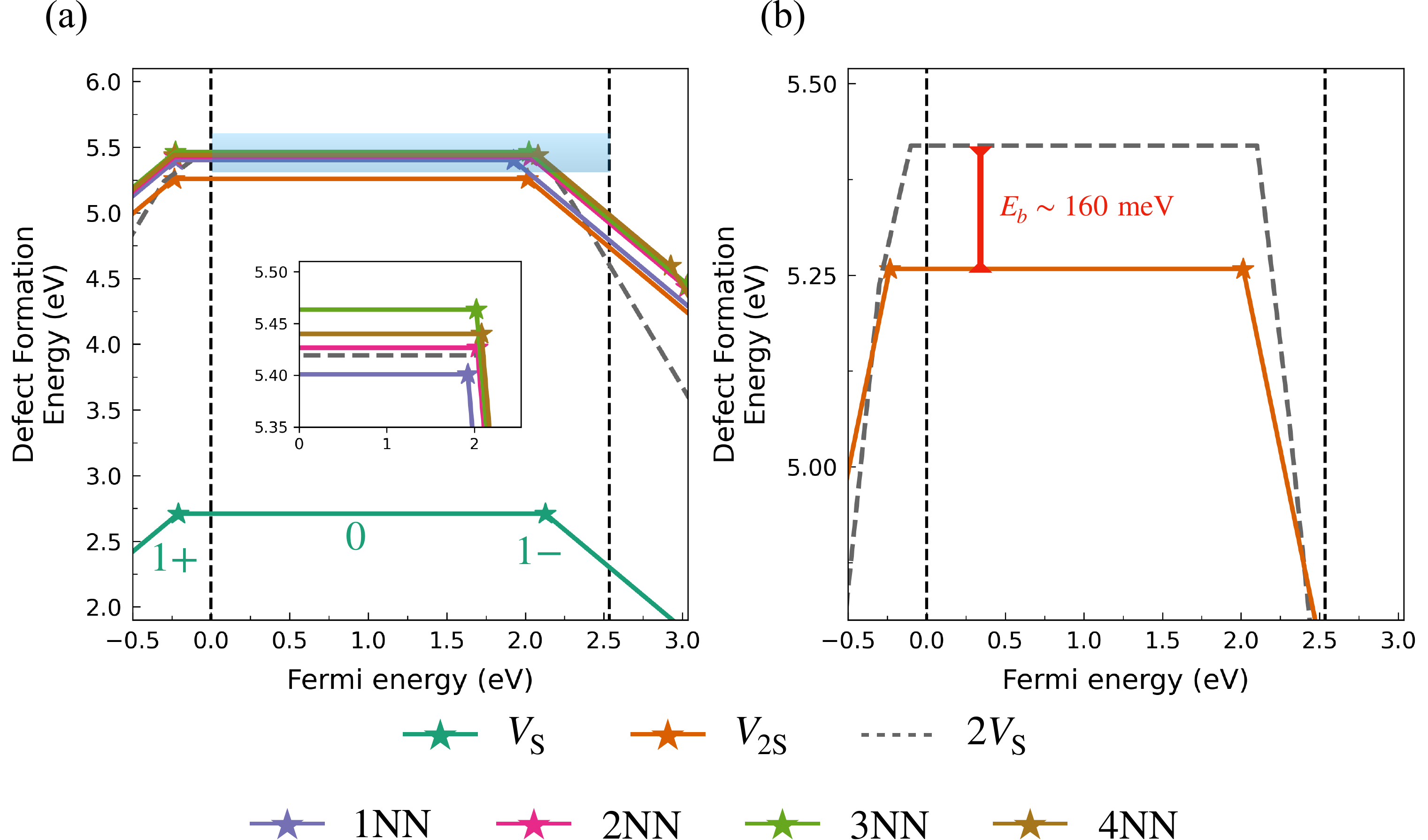}
    \caption{Formation energy of sulfur single vacancy $V_{\mathrm{S}}$, divacancy $V_{\mathrm{2S}}$, and nearest neighbor complexes 1NN to 4NN from charge state 2- to 2+ evaluated under sulfur poor condition (a). Gray dashed line indicating two infinitely separated vacancies $2V_{\mathrm{S}}$ competing with the formation of vacancy paries. Inset shows zoomed-in energy range marked in the shaded area. (b): a binding energy around 160 meV is observed by the divacancy configuration at neutral charge state. Band edge positions are marked with vertical dashed lines. Fermi energy is referenced to the VBM.}
    \label{fig:formation_energy}
\end{figure*}

\begin{figure*}[h!]
    \centering
    \includegraphics[width = 1\textwidth]{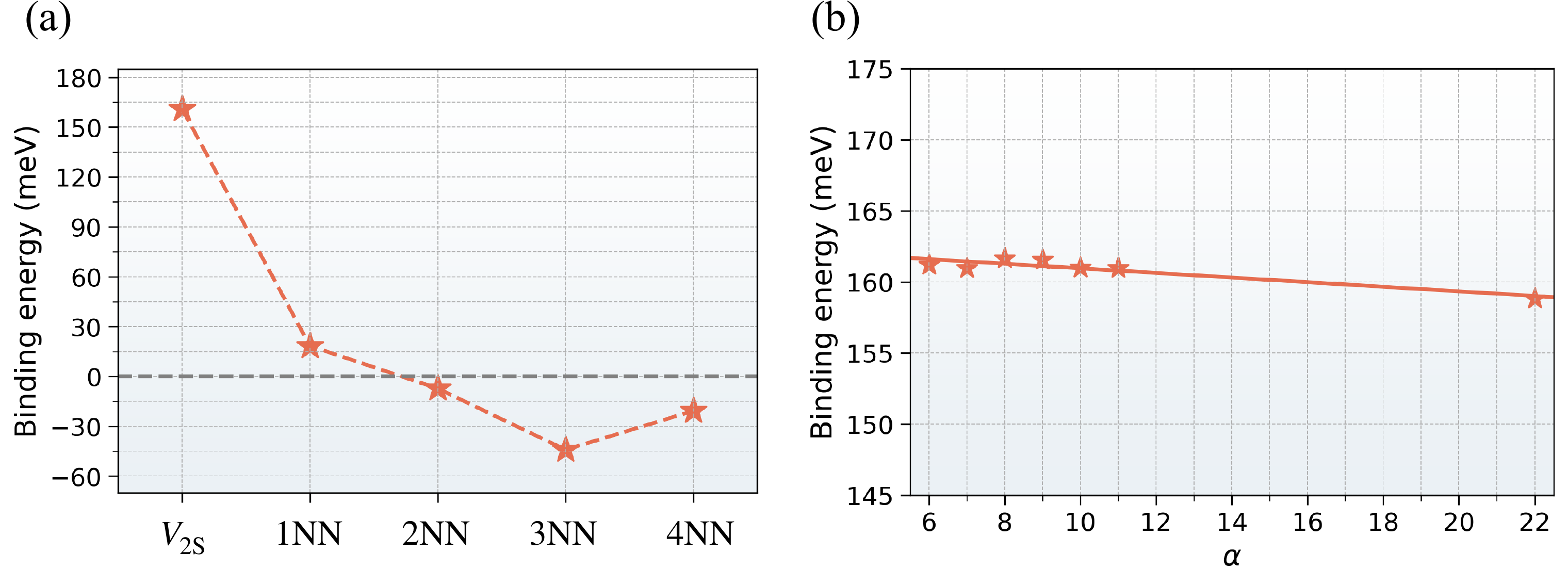}
    \caption{Computed binding energy, dependence on the amount of exact exchange ($\alpha$) employed and the resulted equilibrium constant of divacancy formation. (a): binding energies for divacancy $V_{\mathrm{2S}}$ and nearest neighbor vacancy complexes 1NN through 4NN. (b): linear fit for the divacancy binding energy with different amount of exact exchange.}
    \label{figS:defect_interaction_energy}
\end{figure*}

\newpage
\bibliographystyle{ieeetr} 
\bibliography{main}